\documentclass{pasj00}
\begin{document}
\SetRunningHead{M. Honma et al.}{Astrometry of S269 with VERA}
\Received{2007/03/09}%{yyyy/mm/dd}
\Accepted{2007/05/20}%{yyyy/mm/dd}

%\title{Parallax Measurement of Galactic Star Forming Region Sharpless-269 with VERA: initial results from 1-year monitoring}

%\title{First Parallax Measurement beyond 5 kpc and Constraints on the Outer Rotation Curve of the Galaxy}

%\title{Outer rotation curve of the Galaxy constrained by accurate astrometry with VERA}

\title{Astrometry of Galactic Star Forming Region Sharpless 269 with VERA : Parallax Measurements and Constraint on Outer Rotation Curve}

%%% begin:list of authors

\author{
Mareki \textsc{Honma},\altaffilmark{1,2}
Takeshi \textsc{Bushimata},\altaffilmark{1,3}
Yoon Kyung \textsc{Choi},\altaffilmark{1,4}
Tomoya \textsc{Hirota},\altaffilmark{1}
Hiroshi \textsc{Imai},\altaffilmark{5}\\
Kenzaburo \textsc{Iwadate},\altaffilmark{6}
Takaaki \textsc{Jike},\altaffilmark{6}
Osamu \textsc{Kameya},\altaffilmark{2,6}
Ryuichi \textsc{Kamohara},\altaffilmark{1}
Yukitoshi \textsc{Kan-ya},\altaffilmark{1,7}\\
Noriyuki \textsc{Kawaguchi},\altaffilmark{1,2}
Masachika \textsc{Kijima},\altaffilmark{1,2}
Hideyuki \textsc{Kobayashi},\altaffilmark{1,3,4,6}
Seisuke \textsc{Kuji},\altaffilmark{6}\\
Tomoharu \textsc{Kurayama},\altaffilmark{1}
Seiji \textsc{Manabe},\altaffilmark{2,6}
Takeshi \textsc{Miyaji},\altaffilmark{1,3}
Takumi \textsc{Nagayama},\altaffilmark{5}\\
Akiharu \textsc{Nakagawa},\altaffilmark{5}
Chung Sik \textsc{Oh},\altaffilmark{1,4}
Toshihiro \textsc{Omodaka},\altaffilmark{5}
Tomoaki \textsc{Oyama},\altaffilmark{1}
Satoshi \textsc{Sakai},\altaffilmark{6}\\
Katsuhisa \textsc{Sato},\altaffilmark{6}
Tetsuo \textsc{Sasao},\altaffilmark{8,9}
Katsunori M. \textsc{Shibata},\altaffilmark{1,3}
Motonobu \textsc{Shintani},\altaffilmark{5}\\
Hiroshi \textsc{Suda},\altaffilmark{4,6}
Yoshiaki \textsc{Tamura},\altaffilmark{2,6}
Miyuki \textsc{Tsushima},\altaffilmark{5} and 
Kazuyoshi \textsc{Yamashita}\altaffilmark{1,2}
}

\altaffiltext{1}{Mizusawa VERA Observatory, NAOJ, Mitaka, Tokyo 181-8588}
\altaffiltext{2}{Graduate University for Advanced Studies, Mitaka, Tokyo 181-8588}
\altaffiltext{3}{Space VLBI Project, NAOJ, Mitaka, Tokyo 181-8588}
\altaffiltext{4}{Department of Astronomy, University of Tokyo, Bunkyo, Tokyo 113-8654}
\altaffiltext{5}{Faculty of Science, Kagoshima University, Korimoto, Kagoshima, Kagoshima 890-0065}
\altaffiltext{6}{Mizusawa VERA observatory, NAOJ, Mizusawa, 023-0861, JAPAN}
\altaffiltext{7}{Department of Astronomy, Yonsei University, Seoul, 120-749, Republic of Korea}
\altaffiltext{8}{Ajou University, Suwon 442-749, Republic of Korea}
\altaffiltext{9}{Korean VLBI Network, KASI, Seoul, 120-749, Republic of Korea}

\email{honmamr@cc.nao.ac.jp}

%% `\KeyWords{}' always has to be placed before `\maketitle'.

\maketitle

\begin{abstract}
We have performed high-precision astrometry of H$_2$O maser sources in Galactic star forming region Sharpless 269 (S269) with VERA.
We have successfully detected a trigonometric parallax of $189 \pm 8$ $\mu$as, corresponding to the source distance of $5.28^{+0.24}_{-0.22}$ kpc.
This is the smallest parallax ever measured, and the first one detected beyond 5 kpc.
The source distance as well as proper motions are used to constrain the outer rotation curve of the Galaxy, demonstrating that the difference of rotation velocities at the Sun and at S269 (which is 13.1 kpc away from the Galaxy's center) is less than 3 \%.
This gives the strongest constraint on the flatness of the outer rotation curve and provides a direct confirmation on the existence of large amount of dark matter in the Galaxy's outer disk.

\end{abstract}

\section{Introduction}

Rotation curves, which are the plots of rotation velocities as functions of distance from galaxy centers, are important tools to study the mass distributions in disk galaxies.
Assuming the centrifugal force balances with gravity, one can determine the mass distribution in a galaxy using the rotation curve.
Observations of rotation curves in external galaxies revealed that they are basically flat within (and often beyond) optical disks of spiral galaxies (e.g., Rubin et al. 1980, Rubin 1983, Sofue \& Rubin 2001).
Flat rotation curves observed in external galaxies provided strong evidences for the dark matter in the galaxies' outer regions (e.g., Kent 1986, 1987).

In contrast, the rotation curve of the Milky Way Galaxy remains highly
uncertain particularly in the outer region, although a large number of
efforts have been made to determine it (e.g, Clemens 1985, Merrifield 1992, Brand \& Blitz 1993, Honma \& Sofue 1997, and references therein).
The reason for this is two fold: 1) it has been difficult to measure
accurate distances of the Galactic objects which are used to trace the
rotation curve, such as OB stars and molecular clouds, and 2) in most
cases it has been difficult to measure proper motions of Galactic
objects, and thus only radial velocity (out of the three components of
spatial velocity) could be used to constrain the Galactic rotation.
Therefore, outer rotation curve and hence dark matter distribution in the Galaxy's disk is still highly uncertain, although it is widely believed that the Galaxy's outer rotation curve is flat just like those of extra-galaxies.

In order to determine a precise rotation curve of the Galaxy, high-precision astrometry is essential.
By measuring the accurate position of a star, one can determine a trigonometric parallax $\pi$ and thus source distance as $D=1/\pi$.
In addition, high-precision astrometry also allows us to determine
source proper motions (source motions projected onto the sky plane), and
thus the 3-dimensional space velocity of the source.
Yet accurate astrometric measurements have been done only for sources that are located fairly close to the Sun compared to the Galaxy's size.
For instance, the HIPPARCOS mission (Perryman et al. 1997), the most modern satellite dedicated to optical astrometry, has reached distances of $\sim$300 pc by parallax measurements, which is much smaller than the size of the Galaxy (e.g., $\sim$15 kpc for the radius of the Galaxy's disk).
Hence, at present astrometry of the Galaxy still remains an unexplored issue.

Over the next decades, there will be new missions for astrometry of the Galaxy that aim at 10 $\mu$arcsec accuracy (e.g., SIM\footnote{SIM: http://planetquest.jpl.nasa.gov/SIM/sim\_index.cfm}, GAIA\footnote{GAIA: http://sci.esa.int/science-e/www/area/index.cfm?\\fareaid=26}, JASMINE\footnote{JASMINE: http://www.jasmine-galaxy.org/index.html}).
These are satellite-type missions with optical/infra-red telescopes
orbiting the Earth, where the observations can be made free from the
disturbance of the atmosphere.
At radio wavelengths, high precision astrometry have been performed with ground-based VLBI (Very Long Baseline Interferometry), a radio interferometer with telescopes separated by $\sim 1000$ to $\sim 10000$ km.
The advantage of VLBI is that it provides the highest angular resolution among the existing telescopes at any wavelength.
While normal VLBI observations directly suffer from the fluctuation of atmosphere (mainly due to the water vapor content in the troposphere), a noble way, which is called {`}phase-referencing{'}, has been developed to cancel out the tropospheric fluctuations.
In phase-referencing observations, a few sources (one target and one or
more reference sources) are observed at nearly same time by rapidly
switching the telescope, and then relative positions of the target with
respect to the reference sources can be measured after correcting for the influence of troposphere.
In fact, recently the distance of the Galactic star forming region W3OH
was determined with VLBA (Xu et al. 2006, Hachisuka et al. 2006) to
solve the long-standing ambiguity of the Perseus arm distance by measuring the source distance at 2 kpc.
It has also been shown that the maser emissions from late-type stars (such as Mira variables) can be used to perform kpc-scale astrometry with VLBI (Kurayama et al. 2005).
These results strongly demonstrate the promising future of ground-based VLBI for the Galaxy-scale astrometry.

VERA (VLBI Exploration of Radio Astrometry) is a new Japanese VLBI array dedicated to VLBI astrometry (Honma et al. 2000, Kobayashi et al. 2003).
A unique feature of the VERA telescopes is the dual-beam receiving system, which allows us to observe simultaneously a target and a reference source within 2.2 degree.
This dual-beam system enables to cancel out tropospheric fluctuations much more effectively than switching observations with normal, single-beam telescopes (Honma et al. 2003).
With a target astrometric accuracy of 10 $\mu$arcsec, VERA aims to precisely locate hundreds of maser sources in the Galaxy and to explore its 3D structure and dynamics.
The VERA array was constructed by 2002, and regular observations have been started in the fall of 2003.
Here we present initial results of high precision astrometry with VERA
toward the Galactic star forming regions S269, which is located in the outer regions ($l=196.45^\circ$).
We report on the determination of the smallest parallax in human
history.
Our results provide the strong constraint on the flatness of rotation curve in the outer Galaxy.

\section{Observations and Reductions}

We observed H$_2$O masers in Galactic star forming region Sharpless 269 (S269) with VERA since November of 2004 and here we present the data of 6 epochs that were obtained with the full 4-station array (Mizusawa, Iriki, Ogasawara, and Ishigaki-jima) under relatively good conditions.
The epochs are day of year (DOY) 323 in 2004, DOY 026, 073, 134, 266 and 326 in 2005 (Nov. 18 in 2004, Jan. 26, Mar. 14, May. 14, Sep. 23 and Nov. 21 in 2005), spanning $\sim$1 year.
At each epoch, H$_2$O 6$_{16}$-5$_{23}$ maser line at a rest frequency
of 22.235080 GHz in S269 and a position reference source J0613+1306 were simultaneously observed in dual-beam mode for nearly 9 hours.
Typical on-source integration time was 5 hours for both the target maser and the reference.
The reference source, J0613+1306, is one of the ICRF sources (Ma et
al. 1998) with a correlated flux density of $\sim$300 mJy.
The separation angle between the maser and reference sources is 0.73 degree.
Left-hand circular polarization signals were received for both S269 and J0613+1306, and digitally recorded onto magnetic tapes with the VERA-terminal system at the total data rate of 1024 Mbps.
With 2-bit quantization, this data rate provides a total bandwidth of 256 MHz.
The signals from the two sources are filtered with VERA Digital Filter
(Iguchi et al. 2005) to obtain 1 IF (Intermediate Frequency) channel of
16 MHz for the S269 maser line and 15 IF channels of 16 MHz (240 MHz in total) for J0613+1306.
Correlation processings were made with the Mitaka FX correlator.
For the reference, which is a continuum source, the spectral resolution was 64 points per each 16 MHz channel, which corresponds to a velocity resolution of 3.4 km s$^{-1}$.
For the maser source, frequency and velocity resolutions were 15.625 kHz and 0.21 km s$^{-1}$, respectively.

Since the correlator's a priori delay model is not accurate enough for high precision astrometry, recalculations of precise delay were made after the correlation, and correlated visibilities were corrected for the difference between the first (rather crude) a priori model and second (more accurate) delay model.
The delay recalculation code is based on the geodynamics models described in IERS convention 1996 (McCarthy 1996), and earth orientation parameters (EOP) were taken from IERS bulletin B final values\footnote{http://hpiers.obspm.fr/eop-pc/}, which currently provides the best estimates.
Also, ionospheric delays were corrected based on the global ionosphere
map (GIM), which are produced by the University of Bern every day\footnote{http://www.aiub.unibe.ch/ionosphere.html}.

In the data analysis of each epoch, at first fringes were searched for the reference source J0613+130.
With 240 MHz bandwidth and typical system noise temperature of ~200 K, the reference source, having a flux of $\sim$300 mJy, was easily detected within 1 minute integration, which is much shorter than typical coherence time at 22 GHz (2 to 3 minutes). 
Since its position is accurately known, the fringe parameters of J0613+1306 are also used to calibrate clock offset parameters (such as delay and delay-rate offset).
The phase solutions for J0613+1306 were converted into the phase at
observed maser frequency, and applied to the visibilities of S269
together with the dual-beam phase calibration data.
This dual-beam phase calibration data was taken real-time during the
observations and are based on the correlation of artificial noise sources injected into two beams at each station (Kawaguchi et al. 2000).
After those calibrations, the visibilities of the S269 masers were Fourier transformed to synthesize images, and the positions of the brightness peaks were determined with respect to the reference spot.
In some epochs (especially in summer), the qualities of phase-referenced maps were not high due to residuals in tropospheric delay.
To calibrate them, residual zenith delays were estimated as a constant offset that maximizes the coherence of the phase-referenced map.
Typical residuals of zenith delay are 1 to 5 cm, but in the worst case (during the summer at Ishigaki-jima station) it was as large as 20 cm.

\section{Results}

\begin{figure}[t]
\begin{center}
       \FigureFile(85mm,85mm){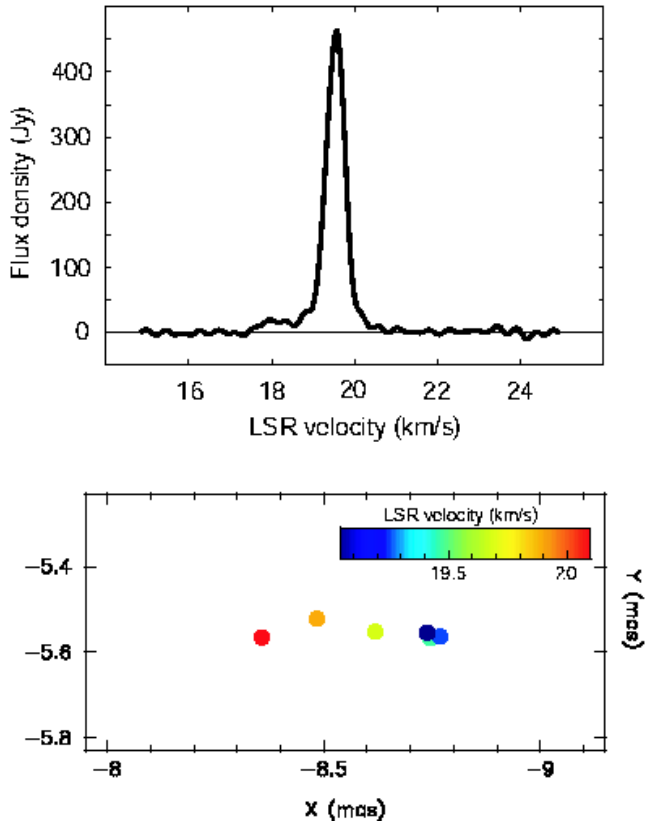}
\end{center}
\caption{ (a, top) Auto-correlation spectrum of S 269 H$_2$O maser emissions taken on DOY 073 in 2005.
(b, bottom) maser spot distribution obtained with VERA on DOY 073 in 2005.
The coordinates are with respect to the tracking center position, which is (06h14m37.08s, +13d49'36.7") in J2000.}
\end{figure}
\begin{table*}
\begin{center}
\caption{The best fit values of parallax $\pi$ and proper motions $\mu_X$ and $\mu_Y$ for the three brightest spots.
The last row gives the values obtained by weighted mean.
Note that positions $X$ and $Y$ are those at the first epoch (DOY 323 in 2004) with respect to the tracking center position of S269, which was taken to be (06h14m37.08s, +13d49'36.7") in J2000.}

\begin{tabular}{cccccc}
$X$ (mas) & $Y$ (mas) & $V_{\rm LSR}$ & $\pi$ (mas) & $\mu_X$ (mas/yr) & $\mu_Y$  (mas/yr)\\
\hline
-8.183 & -5.571  & 19.8 & 0.208$\pm$0.030 & $-0.425\pm$0.032 & $-0.123\pm$0.076 \\
-8.365 & -5.591 & 19.6 & 0.199$\pm$0.012 & $-0.388\pm$0.014 & $-0.118\pm$0.071 \\
-8.427 & -5.602 & 19.3 & 0.176$\pm$0.012 & $-0.457\pm$0.015 & $-0.123\pm$0.072 \\
\hline
mean &          &      & 0.189$\pm$0.008 & $-0.422\pm$0.010 & $-0.121\pm$0.042 \\
\end{tabular}
\end{center}
\end{table*}

The total-power spectrum of S269 taken on DOY 073 in 2005 is shown in figure 1(a).
Basically it consists of a single feature at $V_{\rm LSR}$\footnote{the LSR velocity described here is in traditional definition using the Solar motion of (U, V, W)=(10, 15.4, 7.8) km s$^{-1}$(Kerr \& Lynden-Bell 1986) for comparisons with previous observations.} of $\sim 19.6$ km s$^{-1}$ with a peak intensity of 480 Jy, being consistent with previous single-dish monitoring study (Lekht 2000).
This main feature was always bright and observable for all the epochs presented here.
Figure 1(b) is the maser spot map of the main feature around $V_{\rm LSR}$ of $\sim 19.6$ km s$^{-1}$ (for DOY 2006/073).
Six maser spots were detected in the velocity range from 19.0 to 20.1 km s$^{-1}$, and these maser spots are aligned in the east-west direction on a scale of 0.4 mas.
It is remarkable that the thickness of the feature (spots distribution in the north-south direction) is $\sim 50$ $\mu$as, 10 times smaller than the width in the east-west direction.
The maser distribution also shows a velocity gradient from the east to
the west.
This kind of structure is rather unusual for H$_2$O masers in star
forming regions (which mostly show bipolar structure with discrete 
blue/red-shifted clusters of maser spots), but instead, similar to those
of CH$_3$OH (methanol) masers at 6.7 GHz and 12.2 GHz in terms of
the spot distribution as well as the velocity width (e.g., Minner et al. 2000).
Note that positions in figure 1(b) are the residuals to the tracking center positions of the maser and reference sources, which are taken to be ($\alpha$, $\delta$)=(06h14m37.08s, +13d49'36.7") for S269, and (06h13m57.692764s, +13d06'45.40116") for J0613+1306, both in the J2000 coordinates.
Thus, the absolute position of the brightest $19.6$ km s$^{-1}$ spot at DOY 073 of 2005 is obtained as (06h14m37.07933s, +13d49'36.6945\") with an uncertainty of 1 mas, which mainly comes from the uncertainty of absolute position of the reference source J0613+1306.
The absolute position of the maser feature shown in figure 1(b) agrees well with the position of S269 IRS2w, which is the most luminous infrared source in the S269 regions (Jiang et al. 2003).

\begin{figure}[th]
\begin{center}
       \FigureFile(85mm,85mm){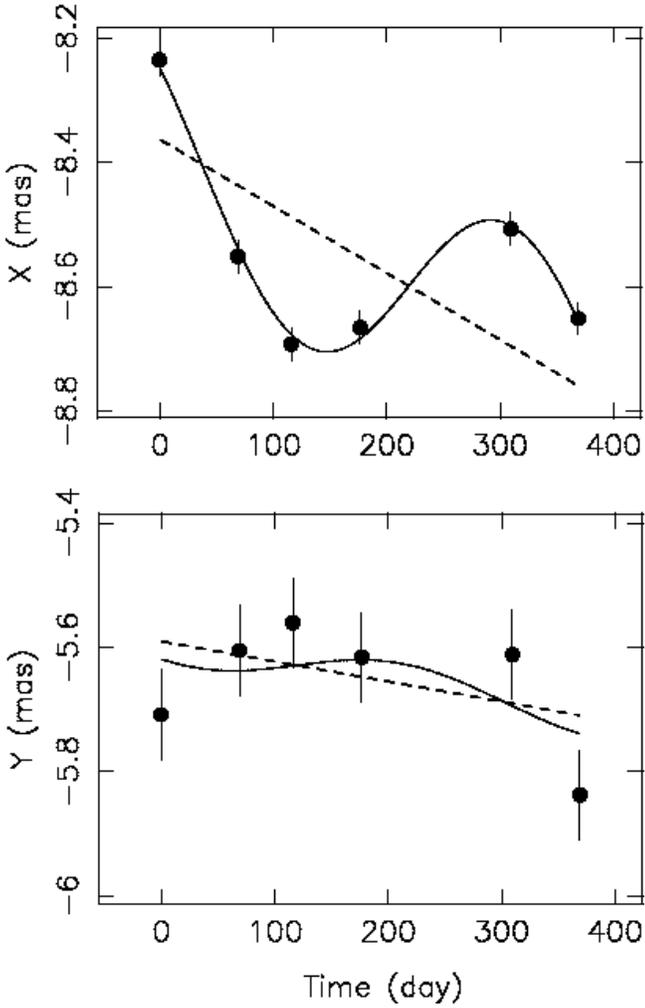}
\end{center}
\caption{Positional variations of the brightest maser spot at $V_{\rm LSR}$=19.6 km $^{-1}$.
Top is for $X$ (east-west) direction and bottom is for $Y$ (north-south) direction.
The time origin is the first epoch of our observations, which is DOY 323 in 2004.
Solid curves are best-fit results with parallax and proper motions, and dashed lines show the best-fit proper motions.}
\end{figure}

Figures 2 show the positional variations of the brightest maser spot in $X$ ($\equiv \cos \delta \Delta \alpha$, east-west offset) and $Y$ ($\equiv \Delta \delta$, north-south offset) directions for a monitoring span of 1 yr.
As clearly seen from the plot (especially in $X$ direction), the positions show systematic sinusoidal modulation with a period of 1 yr.
The phase of observed sinusoidal curve (i.e., the peak date) perfectly matches with that of the expected parallax curve for S269, ensuring that the modulation certainly originates from the parallax of S269.
In figure 2, error bars were estimated as the standard deviation from the best fit with parallax plus linear proper motions.
This error estimate was made because it is difficult to predict the observational error in VLBI astrometry: the error depends on many factors such as residual phase in phase-referencing, error in zenith delay of troposphere and ionosphere, and error in calibration of instrumental offset and so on, and is hardly predictable.
The estimated error bars are 25 $\mu$as for $X$ and 75 $\mu$as for $Y$.
We note that error in $Y$ is three times larger than that in $X$.
This can be explained if one assume that majority of the error comes from the uncertainty in the tropospheric zenith delay, as the tropospheric delay changes apparent elevations of the source.
Detailed consideration on the error is discussed in the next section.
For our parallax measurements, in this paper we consider only the $X$
component because the $Y$-direction error is large and also because S269
is near the ecliptic and the parallax ellipse is highly elongated in the
$X$ direction, making the contribution of $Y$ to the parallax determination smaller.
Here we use the brightest three spots at the radial velocity from $19.4$ to $19.8$ km s$^{-1}$ including the brightest one at $19.6$ km s$^{-1}$ to ensure the high signal-to-noise ratio: as clearly seen from figure 1, the H$_2$O maser spectrum is sharply peaked at $19.6$ km s$^{-1}$, and maser intensity becomes weak at off-peak radial velocities.
Least-squares fits were made to positions of the three maser spot, with parallax $\pi$ as well as proper motions $\mu_X$ (east-west direction) and $\mu_Y$ (north-south direction).
Table 1 summarizes the best fit results.
As seen in table 1, independent analyses for the three spots give
results consistent with each other, from 176 $\mu$as to 208 $\mu$as.
To obtain the best estimate of the parallax $\pi$, we took a weighted mean of the parallaxes, yielding the parallax of S 269 as 189 $\pm$ 8 $\mu$as, where the error bar is estimated as $\sigma^2=1/\sum (1/\sigma^2_i)$.
This corresponds to a source distance of $5.28^{+0.24}_{-0.22}$ kpc.
This is the smallest parallax ever measured to date, demonstrating the high capability of VERA to perform Galactic-scale astrometry.
The distance to S269 is found to be slightly larger than previous estimates, which claimed a distance of $\sim$4 kpc (Moffat et al. 1979).

\begin{figure*}[t]
\begin{center}
	       \FigureFile(120mm,120mm){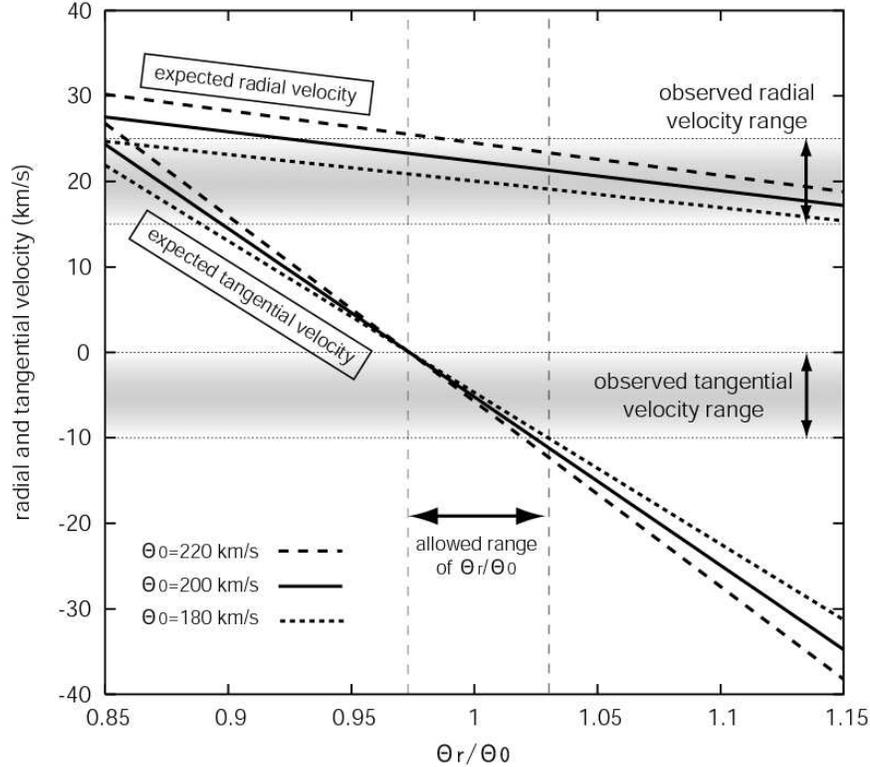}
\end{center}
\vspace{-5mm}
\caption{Plot of radial velocity $v_r$ and tangential velocity $v_l$ with respect to LSR as a function of the rotation velocity ratio $\theta=\Theta/\Theta_0$.
Observed value of $v_l$(=$-5$ $\pm$ $5$) is consistent with the rotation velocity ratio of $\theta=1.00\pm0.03$, showing the rotation velocity is the same at the position of S269 and at the Sun.
Note that $v_r$(=$20$ $\pm$ $5$) can be explained consistently with $\theta=1.00\pm0.03$.}
\end{figure*}

From the fitting results in table 1, the weighted means of the proper motions are obtained as ($\mu_X=-0.422 \pm 0.010$, $\mu_Y=-0.121 \pm 0.042$) mas yr$^{-1}$, respectively.
To convert these observed (heliocentric) proper motions to the ones with respect to LSR, we use the solar motion based on the HIPPARCOS satellite data (Dehnen \& Binney 1998), which is ($U$, $V$, $W$)=(10.0, 5.25, 7.17) km s$^{-1}$.
Using the galactic coordinates of S269 ($l$, $b$)=(196$^\circ$.45, $-1^\circ.96$) and the Galactic plane's position angle of 151.51$^\circ$ there, one can calculate the proper motion projected to the direction of $l$ and $b$ as ($\mu_l$, $\mu_b$)=($-0.184\pm 0.032$, $-0.149\pm 0.029$) mas yr$^{-1}$.
Given the source distance of 5.28 kpc, these proper motions correspond to the velocity vector of ($v_l$, $v_b$)=($-4.60\pm 0.81$, $-3.72\pm 0.72$) km s$^{-1}$, respectively.
These velocity components are remarkably small compared to the rotation speed of the Galaxy, which is an order of $\sim 200$ km s$^{-1}$.
Given that S269 is located in the anti-center region, the small value of $v_l$ indicates that the Galactic rotation velocities at the Sun and at S269 are close to each other and proper motions were cancelled out in our relative proper motion measurements.
Detailed discussion on the Galactic rotation velocity will be done in next section.

\section{Discussion}

\subsection{Sources of Astrometric Error}

As described in the previous section, the astrometric errors estimated from the fitting deviations are 25 $\mu$as for $X$ and 75 $\mu$as for $Y$.
This can be explained if the dominant error source is the uncertainty in the tropospheric zenith delay.
For instance, if we take a typical uncertainty of 3 cm for the the tropospheric zenith delay, then it causes an path length difference of 0.4 mm (=30 mm $\times$ 0.7 deg $/$ 57.3 deg/rad, where 0.7 deg is the separation angle between S269 and the reference source) between the two sources.
This uncertainty in the path length difference roughly corresponds to 40 $\mu$as (=0.4 mm / 2.3$\times 10^9$ mm, where 2.3 $\times 10^9$ mm is the maximum baseline length of the VERA array).
In practice, observations are not done toward the zenith, but sources are at lower elevation angle ($EL$).
As our observations were usually done for elevation angle larger than 20$^\circ$, the effect of zenith delay error is multiplied by a factor of 1 (corresponding to $EL=90^\circ$) to $\sim$3 ($EL=20^\circ$) depending on the source elevation.
If this factor is taken into account, one can expect an astrometric error of 40 to 120 $\mu$as depending the EL distribution, and the astrometric error in $Y$ obtained above (75 $\mu$as) is certainly in this possible range.
On the other hand, the astrometric error in the $X$ direction can be
suppressed for two reasons: first, the source pair considered here has a
smaller separation in $X$ direction than in $Y$ direction, and second, the observational track in each epoch is roughly symmetric with respect to the meridian transit (i.e., each epoch has nearly same track before and after the transit), and this symmetry can help to reduce the astrometric error in the $X$ direction caused by the tropospheric zenith delay offset.

\begin{figure*}[t]
\begin{center}
       \FigureFile(130mm,130mm){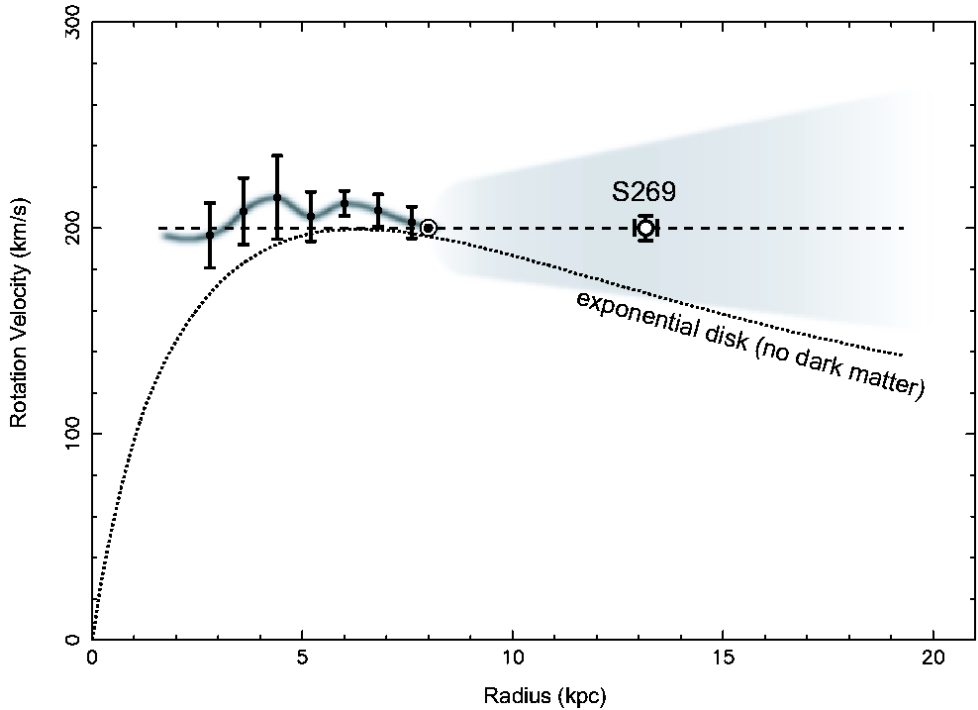}
\end{center}
\caption{Rotation curve of the Milky Way Galaxy obtained in previous
 studies together with our result on S269 using Galactic constants of $R_0$=8 kpc and $\Theta_0$=200 km $^{-1}$.
Dashed line is the flat rotation curve with $\Theta$=200 km $^{-1}$,
 shadowed area shows possible range of outer rotation curves considered in previous studies (Honma \& Sofue 1997).
Points at $R\le 8$ kpc are inner rotation curves determined from the tangential velocities of Galactic HI gases (Honma \& Sofue 1997), with a smoothed fit (thick curve).
Dotted curve is the rotation curve for an exponential disk, corresponding to a constant mass-to-light ratio disk without dark matter.
Discrepancy between the observed point for S269 and the exponential disk is evident, demonstrating the existence of large amount of dark matter in the outer region of the Galaxy.}
\end{figure*}

Other possible sources of astrometric error are those in station positions, delay model, and ionosphere.
Currently, VERA station positions are determined with an accuracy of $\sim$ 3 mm based on geodetic observations at S/X (2/8 GHz) bands carried out every two weeks.
Also, our delay calculation code is compared with
CALC\footnote{http://gemini.gsfc.nasa.gov/solve/} developed by the NASA/GSFC VLBI group, which is the international standard of delay model, and it turned out that the difference between the two codes is less than 2 mm.
Therefore, these errors are smaller than that of zenith delay by an order of magnitude.
Regarding the ionosphere, its contribution is small at 22 GHz, and corrections with GPS-based Global Ionosphere Map (GIM), provided by University of Bern, are precise enough for 10 $\mu$as astrometry.
We note that the trend of the larger error in $Y$ direction was also found in other observations of VERA.
Therefore, the larger error in $Y$ is not a special phenomenon of only the S269 observations but rather common to all of VERA observations, and is most likely to originate from the tropospheric zenith delay error.

\subsection{Constraint on Galactic Rotation}

Here we use the proper motions and parallax obtained in the previous section to constrain the Galactic rotation velocity at the position of S269.
First, the small proper motion perpendicular to the Galactic plane
($v_b=-3.72\pm0.72$ km s$^{-1}$) indicates that S269 basically partakes in galactic rotation and also that the H$_2$O maser proper motions truly reflect the systemic motion of S269.
Radial velocity measurements supports the latter idea: while the maser
emission is peaked at 19.6 km $^{-1}$, the peak velocity of HII region
observed with SII lines is 16.5 km s$^{-1}$, and the systemic velocity
of the associated molecular cloud observed in CO is 17.7 km s$^{-1}$ (Godbout et al. 1997), which agrees well with the maser radial velocity within 3 km s$^{-1}$.
Therefore, from the radial velocity as well as proper motions
perpendicular to the Galactic plane, one can expect that the peculiar
velocity of the maser source with respect to pure Galactic rotation is as small as $\sim$5 km s$^{-1}$.

From this fact one can expect that the proper motion in $l$ direction
($v_l$) basically reflects the difference of the galactic rotation
velocities at the position of S269 and at the Sun, and also that one can constrain the outer rotation velocity at the position of S269.
In fact, S269 is located near the galactic anti-center region ($l=196^\circ.45$), and thus the lack of a large proper motion in $l$ direction indicates that the galactic rotation speed there is close to that at the Sun.
If a source is on perfect Galactic rotation, radial and tangential velocities with respect to LSR observers can be written as,
\begin{equation}
\label{eq:V_r}
v_r = \left(\frac{\Theta}{R}-\frac{\Theta_0}{R_0}\right) R_0 \sin l,
\end{equation}
\begin{equation}
\label{eq:V_t}
v_l = \left( \frac{\Theta}{R}-\frac{\Theta_0}{R_0}\right) R_0 \cos l - \frac{\Theta}{R} D,
\end{equation}
For the component perpendicular to the Galaxy plane, obviously $v_b=0$.
These equations relate the observed velocities to the rotation velocity at the source ($\Theta$) through the Galactic constants $R_0$ and $\Theta_0$.
Figure 3 shows the plot of expected $v_r$ and $v_l$ for S269 as a function of $\theta\equiv \Theta/\Theta_0$ (rather than $\Theta$ for seeing the difference of the rotation velocities at S269 and the Sun).
Here the Galactic constant $R_0$ is assumed to be 8.0($\pm$0.5) kpc (Reid 1993), and three cases of $\Theta_0$, 180, 200, and 220 km s$^{-1}$ are considered as recent determinations of $\Theta_0$ still vary substantially (Olling \& Merrifield 1998, Miyamoto \& Zhu 1998, Reid \& Brunthaler 2004).
As seen in figure 3, the tangential velocity $v_l$ has a strong dependence on $\theta$, and thus observed $v_l$ can be used to constrain the rotation velocity.
Considering that the velocity component perpendicular to the Galactic plane is 3.7 km s$^{-1}$ and that the radial velocity differences between maser and other lines (such as CO and SII) are also $\sim$3 km s$^{-1}$, here we can safely assume the possible range of the true tangential velocity as $v_l$=-5 $\pm$ 5 km s$^{-1}$ and that of the true radial velocity as $v_r=$20 $\pm$ 5 km s$^{-1}$.\footnote{Note that the radial velocity considered here is slightly different from the traditionally-defined $V_{\rm LSR}$ since the Solar motion considered here is the one recently obtained from HIPPARCOS data (Dehnen \& Binney 1998). However, the difference in $V_{\rm LSR}$ and $v_r$ is not significant, being 3.3 km $^{-1}$ for S269.}
From figure 3, this tangential velocity range is obtained if $\theta$ lies between 0.97 and 1.03.
Therefore, at the position of S269, $\theta$ should be $1.00\pm 0.03$, i.e., the rotation velocity at the S269 ($\Theta$) must be the same with $\Theta_0$ within 3\% level.
This is the strongest constraint of the rotation velocity in the outer galaxy ever obtained.
Note that the same argument is possible based on the radial velocity $v_r $, but the constraint is not as strong as that from the tangential velocity, because the gradient of $v_r(\theta)$ plot is much shallower. 
However, we note that the radial velocity $v_r=20 \pm 5$ km s$^{-1}$ can be consistently explained by the rotation velocity ratio of $\theta\sim 1$.

In previous works, the Galactic rotation curve has an uncertainty up to 100 km s$^{-1}$ in the outer region if one includes the strong dependence on Galactic constants $\Theta_0$ (Honma \& Sofue 1997).
This situation is summarized in figure 4, showing the area of uncertainty in previous studies.
The point for S269 determined in this study is also shown in figure 4.
The coincidence of rotation velocities at the Sun and S269 simply indicates that the rotation curve there is basically flat, as was known for rotation curves of other spiral galaxies (Rubin et al. 1980, Rubin 1983, Sofue \& Rubin 2001).
In disk galaxies like the Galaxy, optical surface brightness obeys an exponential law, i.e., $I(r)=I_0 \exp (-r/h)$, where $r$ is the radius and $h$ is the disk scale length.
Assuming that surface brightness traces the mass density (i.e., constant Mass-to-Light ratio), one can calculate the rotation curve of optical disk assuming no dark matter (Freeman 1970).
In figure 4, we also showed such a rotation curve for the Galaxy's disk, assuming a disk scale length of $h=3$ kpc and a maximum rotation velocity of 200 km $^{-1}$.
As seen from figure 4, such a rotation curve without dark matter was not completely ruled out previously.
However, the rotation velocity measurement for S269 evidently shows the discrepancy between the observed rotation curve and that expected from the optical disk without dark matter, providing a strong confirmation on the existence of dark matter in the Galaxy's outer region.
At the position of S269, the rotation curve of the exponential disk gives $V_{\rm exp}=168$ km s$^{-1}$ while the astrometric measurement of S269 provides $V_{\rm obs}=200$ km s$^{-1}$.
In case of a spherical mass distribution, the enclosed mass $M_r$ within a radius $r$ is proportional to $V^2$, as $M_r$ can be written as $M_R\approx r V^2/G$.
The values of $V_{\rm exp}$ and $V_{\rm obs}$ obtained above give $(V_{\rm exp}/V_{\rm obs})^2 = 0.70$, and thus within the position of S269 at least $\sim 30$ \% of the enclosed mass must be composed of dark matter.

Although this study presents astrometric measurements for only one source, the present study demonstrates the high capability of VERA in studying the Galactic rotation based on Galaxy-scale astrometry.
During next decade, VERA continues astrometric observations of nearly
one thousands Galactic maser sources and will provide an accurate
rotation curve over the whole Galaxy's disk as well as an accurate
description of the dark matter distribution of the Galaxy.

\bigskip

We are grateful to the referee, Prof. Karl Menten, for carefully reading the manuscript.
One of the authors (MH) acknowledges financial support from grant-in-aid (No.16740120) from the Ministry of Education, Culture, Sports, Science and Technology (MEXT).
Authors also would like to thank all the supporting staffs at Mizusawa VERA observatory for helping observations.

%%%%
% References
%%%%
%\clearpage

\end{document}